\documentclass[cits,A4]{PoS}
\usepackage{amsmath,amssymb}
\usepackage{fontenc}
\usepackage{times}
\usepackage{mathptmx}
\usepackage{epsfig}
\usepackage{graphicx}
\usepackage{slashed}

\newcommand{\beq}{\begin{equation}}
\newcommand{\eeq}{\end{equation}}
\newcommand{\bea}{\begin{eqnarray}}
\newcommand{\eea}{\end{eqnarray}}
\newcommand{\ba}{\begin{align}}
\newcommand{\ea}{\end{align}}
\newcommand{\bfig}{\begin{figure}}
\newcommand{\efig}{\end{figure}}

\newcommand{\D}{\displaystyle}

\newcommand{\gev}{\, \text{GeV}}

\newcommand{\tin}{t_{\rm in}}

\newcommand{\tplus}{t_{+}}

\newcommand{\omnes}{{\cal{O}}}

\title{Constraining the properties of $K_{l3}$ form factors with analyticity and unitarity}

\ShortTitle{Constraining the properties of  $K_{l3}$ $\cdots$}

\author{\speaker{Gauhar Abbas}$^a$, B. Ananthanarayan$^a$,
Irinel Caprini$^b$, I. Sentitemsu Imsong$^a$\\
  \llap{$^a$}Centre for High Energy Physics\\
        Indian Institute of Science\\
        Bangalore, India 560012\\\\
  \llap{$^b$}National Institute of Physics and Nuclear Engineering\\
POB MG 6, Bucharest, R-76900, Romania\\
        E-mail: \email{gabbas@cts.iisc.ernet.in}}

\abstract{Bounds on the  shape parameters of $K_{l3}$ form factors 
are investigated in a general model independent framework.  
Using the method of unitarity bounds  for correlators evaluated in pQCD, chiral symmetry and experimental information on the phase 
and modulus of $\pi K$ 
form factors up to an energy $\tin=1 {\rm GeV}^2$, we constrain the shape parameters of the scalar and vector form factors, 
the value of the scalar form factor at $t=M_\pi^2-M_K^2$,  and exclude regions 
where zeros can exist on the real energy line and in the complex energy plane.}

\FullConference{ 10th International Symposium on Radiative Corrections (Applications of Quantum Field Theory to Phenomenology) 
- Radcor2011\\
September 26-30, 2011\\
Mamallapuram, India}

\begin{document}

\section{Introduction}
Strangeness changing leptonic decays of kaons are an important  
test of the standard model.  In particular, the  $K_{l3}$ decays 
\begin{equation}
 K \rightarrow \pi + l + \nu_l,\, l=e,\, \mu
\end{equation}
are a sensitive probe of
QCD at low energies~\cite{Chounet:1971yy}.  In addition,  these decays provide for the most precise 
determination of the 
Cabibbo-Kobayashi-Maskawa (CKM) matrix element $V_{us}$ \cite{Antonelli:2010yf,Cirigliano:2011ny},
crucial for testing 
the unitarity of the CKM matrix.  The dominant source of uncertainty in the extraction of $V_{us}$, resides
in the experimentally determined quantity $|f_+(0) V_{us}|$.  The $K_{\ell3}$ decay rates were measured 
by BNL-E865, KLOE, KTEV, ISTRA+ and NA48,
for a recent review see \cite{Veltri:2011zk}.  
The decay of a kaon to a pion, charged lepton and a neutrino
is described by the matrix element
\begin{equation}
\langle \pi^0(p') | \overline{s}\gamma_\mu u |K^+(p) \rangle  =  \frac{1}{ \sqrt{2}}[(p'+p)_\mu f_+(t)+(p-p')_\mu f_-(t)],
\end{equation}
where $f_+(t)$ is the vector form factor and the combination
\begin{equation}\label{eq:f0}
f_0(t)=f_+(t)+\frac{t}{M_K^2-M_\pi^2} f_-(t)
\end{equation}
is known as the scalar form factor. The matrix element for the charged pion and the neutral kaon 
is related to Eq. (\ref{eq:f0}) by isospin symmetry.
For the scalar form factor, expansion at $t=0$
\begin{equation}
       f_0(t) = f_+(0)\left(1 +\lambda_0'\, \frac{t}{M_\pi^2} +  \frac{1}{2} \lambda_0''\, \frac{t^2}{M_\pi^4} + \cdots\right)
       \label{eq:f0exp},
\end{equation}
defines the slope $\lambda_0'$ and the curvature $\lambda_0''$ parameters. 
Analogously we define the expansion for the vector form factor.  For a  precise determination of  $|V_{us}|$, it is important to 
improve the accuracy of  the parameterizations of the form factors 
using additional theoretical and experimental information.  We use  inputs from current algebra, perturbative QCD, lattice and 
chiral perturbation
theory to provide stringent bounds on the slope and curvature parameters of the parameterizations of the form factors.
More detailed discussions may be found in \cite{Abbas:2010ns,Abbas:2010jc,Abbas:2009dz}.  
\section{Formalism}\label{formalism}
Analyticity is the ideal tool for relating the information 
from the unitarity cut to the semileptonic range.  The formalism applied in this work (see
also \cite{Abbas:2010ns,Abbas:2010jc,Abbas:2009dz}) exploits the fact that a bound 
on an integral involving the modulus squared of the form factors 
along the unitarity cut is known  from the dispersion relation 
satisfied by a certain QCD correlator.   For scalar form factor this reads
\begin{equation}
\label{chi}
\chi_{_0}(Q^2)\equiv \frac{\partial}{ \partial q^2} \left[ q^2\Pi_0 \right]
= \frac{1}{\pi}\int_{t_+}^\infty\!dt\, \frac{t {\rm Im}\Pi_0(t)}{ (t+Q^2)^2} \,,
\end{equation}

\begin{equation}
\label{im2}
{\rm Im} \Pi_0(t) \ge \frac{3}{2} \frac{t_+ t_-}{ 16\pi}
\frac{[(t-t_+)(t-t_-)]^{1/2}}{ t^3} |f_0(t)|^2  \,,
\end{equation}
with $t_\pm=(M_K \pm M_\pi)^2$.
Similar expression, involving the correlator $\chi_1(Q^2)$, can be
written down for  the vector form factor.
We can now use the conformal map $t\to z(t)$
\begin{equation}\label{eq:z}
z(t)=\frac{\sqrt{t_+}-\sqrt{t_+-t}}{\sqrt{t_+}+\sqrt{t_+-t}}\,,
\end{equation}
that maps the cut $t$-plane onto the unit disc $|z|<1$ in the $z\equiv z(t)$ 
plane, with $\tplus$  mapped onto $z = 1$, the point at 
infinity to $z = -1$ and the origin to $z=0$. This mapping
transforms the relations Eqs. (\ref{chi}) and (\ref{im2}) to 
\begin{equation}
\label{eq:I}
{1\over 2 \pi} \int_0^{2\pi} d\theta|g(\exp(i\theta))|^2 \leq I
\end{equation}
where
\begin{equation}\label{eq:gz}
g(z) = f_0 (t(z)) w(z).	
\end{equation}
The function  $w(z)$ is called outer function and can be calculated analytically in our case.
The function $g(z)$ is analytic within 
the unit disc and can be expanded as:
\beq	
g(z)=g_0+g_1 z+ g_2 z^2 + \cdots,
\eeq
and Eq. (\ref{eq:I}) implies
\beq\label{eq:gkI}
\sum_{k=0}^\infty g_k^2 \leq I.
\eeq
Truncating the above series after finite number of terms gives equations for constraints on the shape parameters.
Improvement of the bound results if
$f_0(t)$ is known at a number of real points $t_i$.
Using a Lagrange multiplier method, we get a determinantal inequality which leads to a quadratic form in the shape parameters 
that is bounded by a known quantity.
In order to exploit the knowledge of the phase, the dispersion contribution from $t_+$
to $\tin$ should be subtracted from the pQCD value,
\begin{equation}\label{eq:I1}
I'= \chi_0(Q^2) - \frac{3}{2} \frac{t_+ t_-}{ 16\pi^2} \int_{t_+}^{\tin}\!dt\, \frac{[(t-t_+)(t-t_-)]^{1/2} |f_0(t)|^2}{t^2 (t+Q^2)^2}\,,
\end{equation}
which requires the knowledge of $|f_0{(t)}|^2$ in the region $t_+ \leq t \leq \tin$.
Now Eqs. (\ref{chi}) and (\ref{im2}) can be written in the following form
\beq\label{eq:hI}
\int_{\tin}^\infty {\rm d}t \rho(t) |\omnes(t)|^2 |h(t)|^2 \leq I^\prime
\eeq
for a known weight function
$\rho(t)$, and where the function  $h(t)$ is analytic in the $t$-plane cut for $t>\tin$ and given by
\beq\label{eq:h}
 h(t)= f_0(t) [\omnes(t)]^{-1},
\eeq 
where we implement the phase information by considering the Omn\`es function:
\begin{equation}    \label{eq:omnes}
 \omnes(t) = \exp \left(\D\frac {t} {\pi} \int^{\infty}_{\tplus} dt \D\frac{\delta (t^\prime)} {t^\prime (t^\prime -t)}\right),
\end{equation}
where $\delta(t)$ is the $I=1/2$ elastic S-wave $K\pi$ scattering
phase, in the elastic region and arbitrarily Lipschitz continuous
above $\tin$,  leading to an extended formalism. This is based on the observation first made in \cite{Caprini:1999ws} which
points out that the phase of the Omn\`es function can compensate for that of
the form factor in the region ($t_+,$ $\tin$), thereby delaying the onset of
the branch point to $\tin$.  
The Eq. (\ref{eq:hI}) can be brought into a canonical form by making 
the conformal transformation
\beq\label{eq:ztin}
\tilde z(t) = \frac{\sqrt{\tin}-\sqrt {\tin -t} } {\sqrt {\tin}+\sqrt {\tin -t}}\,,
\eeq
which maps the complex $t$-plane cut for $t>\tin$ onto the unit disk in the $z$-plane defined by $z=\tilde z(t)$. 
Now  Eq. (\ref{eq:hI}) can be written as
\beq\label{eq:gI1}
\frac{1}{2 \pi} \int^{2\pi}_{0} {\rm d} \theta |g(\exp(i \theta))|^2 \leq I^\prime.
\eeq

The function $g(z)$ is now defined by 
\begin{equation}
 g(z) = w(z)\, \omega(z) \,f_0(\tilde t(z)) \,[O(z)]^{-1},
\end{equation}
where the outer
function for the Omn\`es function given by:
\beq\label{eq:omega}
 \omega(z) =  \exp \left(\D\frac {\sqrt {\tin - \tilde t(z)}} {\pi} \int^{\infty}_{\tin} {\rm d}t^\prime \D\frac {\ln |\omnes(t^\prime)|}
 {\sqrt {t^\prime - \tin} (t^\prime -\tilde t(z))} \right).
\eeq 
For more information and the review of the 
formalism, see \cite{Abbas:2010ns,Abbas:2010jc,Abbas:2009dz}. 

\section{Experimental and theoretical information}
We briefly give a  description of different inputs used
for deriving the improved bounds.
\subsection{QCD correlators}
In the limit $Q^2 >>\Lambda^2_{QCD}$, the correlators $\chi_1(Q^2)$ and $\chi_0(Q^2)$ can be calculated by perturbative QCD. 
Recent calculations to order $\alpha_s^4$  \cite{Baikov:2005rw,Baikov:2008jh} give
\begin{equation}
\label{ip}
I_{+}=\chi_1(Q^2)=\frac{1} {8 \pi^2 Q^2}
(1+\frac{\alpha_s} {\pi}-0.062 \alpha_s^2 
-0.162 \alpha_s^3-0.176\alpha_s^4),
\end{equation}
\begin{equation}
\label{im}
I_{0}=\chi_0(Q^2)=\frac{3(m_s-m_u)^2} {8 \pi^2 Q^2}
(1+1.80\alpha_s+4.65 \alpha_s^2  +15.0 \alpha_s^3+57.4\alpha_s^4 ).
\end{equation}
We have omitted the power corrections due to nonzero masses and QCD condensates, as they are negligible.
We get $\chi_1(2 \gev)=(343.8 \pm  51.6 )\times 10^{-5}
\, {\rm GeV}^{-2}$ and  $\chi_0(2 \gev)= (253\pm 68)\times 10^{-6}$.  
\subsection{Low-energy theorems}\label{sec:low}
The symmetries of QCD at low energies are very useful sources of information for our formalism.  The vector form factor becomes equal
to the scalar form factor at $t=0$.  The SU(3) symmetry implies $f_+(0)=1$.  The deviations from this limit are small due to 
Ademollo-Gatto theorem.    
Recent determinations from the lattice give $f_+(0)=0.964(5)$ \cite{Boyle:2007qe}.
In the case of the scalar form factor, 
current algebra relates the value of the scalar form factor  
at the Callan-Treiman (CT) point
$\Delta_{K\pi}\equiv M_K^2-M_\pi^2$ 
to the ratio  $F_K/F_\pi$ of the decay constants \cite{Callan:1966hu,Dashen:1969bh}:
\beq\label{eq:CT1}
f_0(\Delta_{K\pi})=F_K /F_\pi  +\Delta_{CT}.
\eeq 
In isospin limit, $\Delta_{CT}= -3.1\times 10^{-3}$ to one loop \cite{Gasser:1984ux} and
$\Delta_{CT}\simeq 0$ to two-loops in chiral perturbation theory \cite{Kastner:2008ch,Bijnens:2003uy,Bijnens:2007xa}. 
At $\bar{\Delta}_{K\pi}(=-\Delta_{K\pi})$, a
soft-kaon result \cite{Oehme} relates the value
of the scalar form factor to $F_\pi/F_K$ 
\beq\label{eq:CT2}
f_0(-\Delta_{K\pi})=F_\pi/F_K  +\bar{\Delta}_{CT}.
\eeq
Recent lattice evaluations give $F_K/F_\pi=1.193\pm 0.006$ \cite{Lellouch:2009fg,Durr:2010hr}.
A calculation  in ChPT to one-loop in the isospin limit \cite{Gasser:1984ux} gives $\bar{\Delta}_{CT}=0.03$, but the  higher order 
ChPT corrections are expected to be  larger in this case. 
In the present work we use as input the values of the vector and scalar form factor at $t=0$. For the scalar form factor we impose 
also the value $f_0(\Delta_{K\pi})$ at the first CT point.  As discussed in \cite{Abbas:2009dz}, due to the poor 
knowledge of $\bar{\Delta}_{CT}$, the low-energy theorem Eq.(\ref{eq:CT2}) is not useful for  further constraining the shape of 
the $K_{\ell 3}$ form factors at low energies.  On the other hand we obtain bounds on $\bar{\Delta}_{CT}$. 
\subsection{Phase and modulus along the  elastic region of the cut}
As mentioned in Sec.~\ref{formalism}
the bounds can be improved if the phase of the form factor along the elastic part of the unitarity cut is known from an independent
source.  According to the Fermi-Watson theorem, the phase of the form factor coincides with the phase of the scattering amplitude 
along the elastic
part of the unitarity cut.  
In our calculations  we use below $\tin$ the phases  from \cite{Buettiker:2003pp,ElBennich:2009da} for the scalar form factor, 
and from \cite{Moussallam:2007qc, Bernard:2009zm} for the vector form factor.
We recall that, while the standard dispersion approaches  require a choice of the phase above the inelastic threshold $\tin$,  
the present formalism is 
independent of this ambiguity \cite{Abbas:2010jc}.     Above $\tin$ we have taken  $\delta(t)$ as a smooth function 
approaching $\pi$ at high energies. The results are  independent of the choice of the 
phase for  $t>\tin$. 
We have checked numerically this independence with high precision.

To estimate  the low-energy integral in Eq. (\ref{eq:I1}),  we use the  Breit-Wigner parameterizations of  $|f_+(t)|$ and $|f_0(t)|$ 
in terms of the resonances given by the
Belle Collaboration  \cite{Belle} for fitting the rate of  $\tau\to K\pi\nu$ decay. This leads to the value $31.4 \times 
10^{-5}\,{\rm GeV}^{-2}$ for the vector form factor and $60.9 \times 10^{-6}$  
for the scalar form factor.    By combining with the values of $I_{+,0}$ defined in Eqs. (\ref{ip})-(\ref{im}),  we obtain 
\beq\label{eq:Ip}
I_+'=(312 \pm 69) \times 10^{-5} \,
{\rm GeV}^{-2}, \quad I_0'=(192 \pm 90)\times 10^{-6}.
\eeq
\begin{figure}[htb]
\begin{center}\vspace{0.5cm}
\vspace{0.35cm}
  \includegraphics[width=0.45 \textwidth]{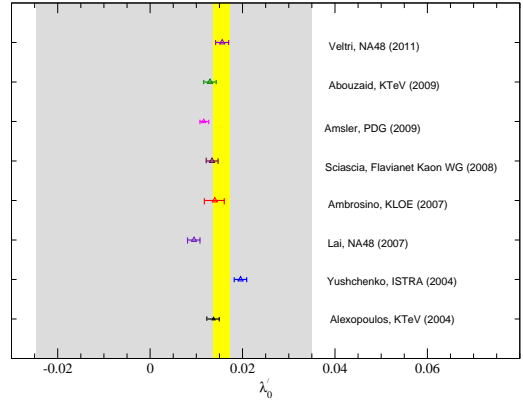}
\caption{The allowed range for the slope of the scalar form factor, when we include phase, modulus and the Callan-Treiman constraint 
(yellow band).  The grey band shows the range  without using the Callan-Treiman constraint.
}
\label{fig1}
\end{center}
\end{figure}
 \section{Results}
\begin{figure}[htb]
\begin{center}
\vspace{0.35cm}
 \includegraphics[width=0.45 \textwidth]{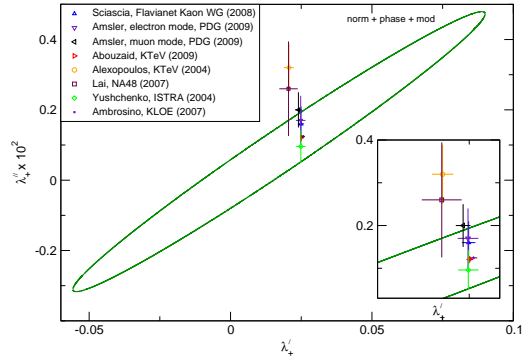}
\caption{The best constraints for the slope and curvature of the vector 
form factor in the slope-curvature plane, where the allowed region is the
interior of the ellipse.
}
	\label{fig2}
\end{center}
\end{figure}
\begin{figure}[ht]
\begin{minipage}[b]{0.5\linewidth}
\centering
\includegraphics[scale=0.3]{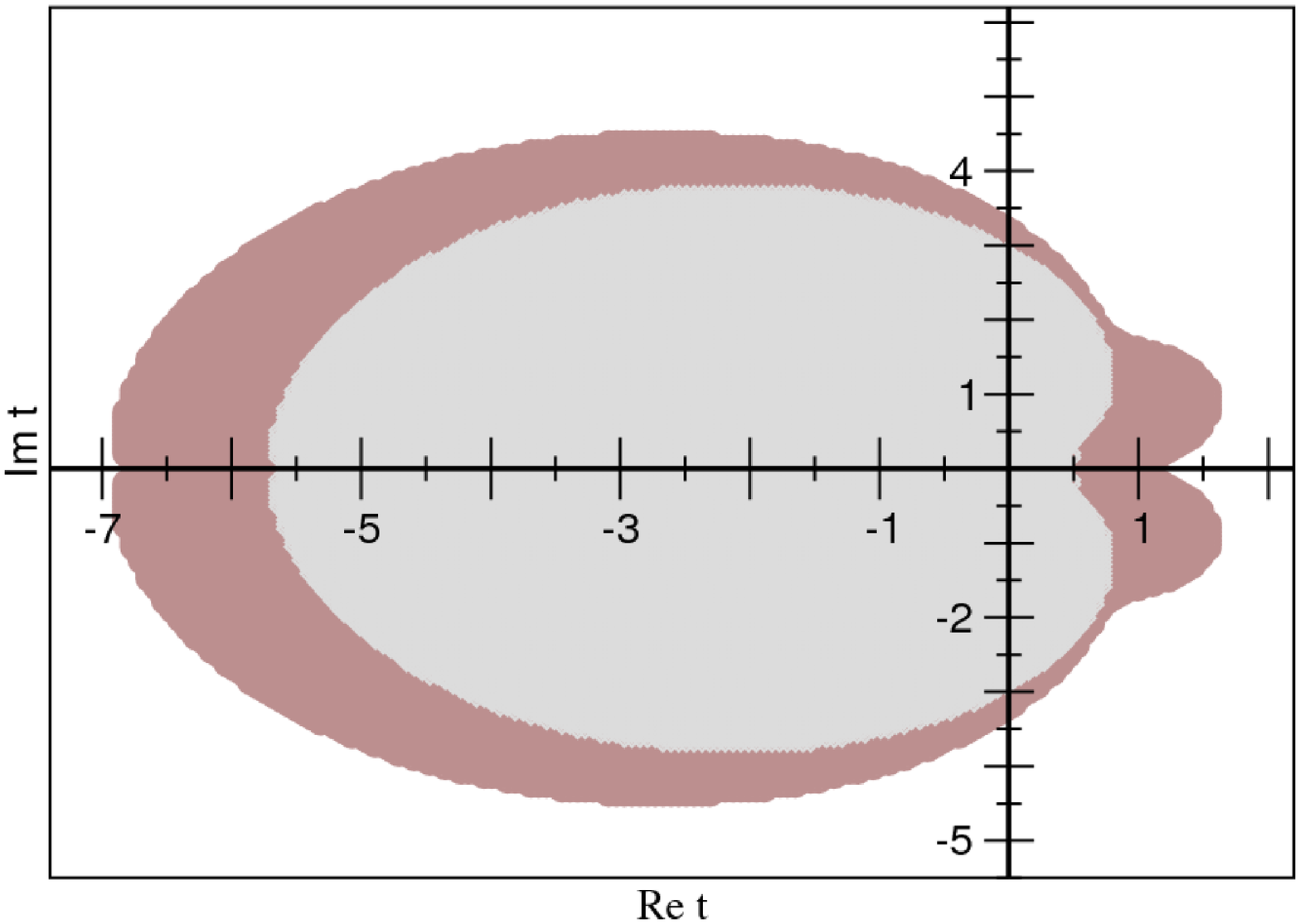}
\caption{Domain without zeros for the scalar form factor: the small domain is obtained 
without including phase and modulus in the elastic region, bigger one using phase, modulus and Callan-Treiman constraint. }
\label{fig3}
\end{minipage}
\hspace{0.5cm}
\begin{minipage}[b]{0.5\linewidth}
\centering
\includegraphics[scale=0.3]{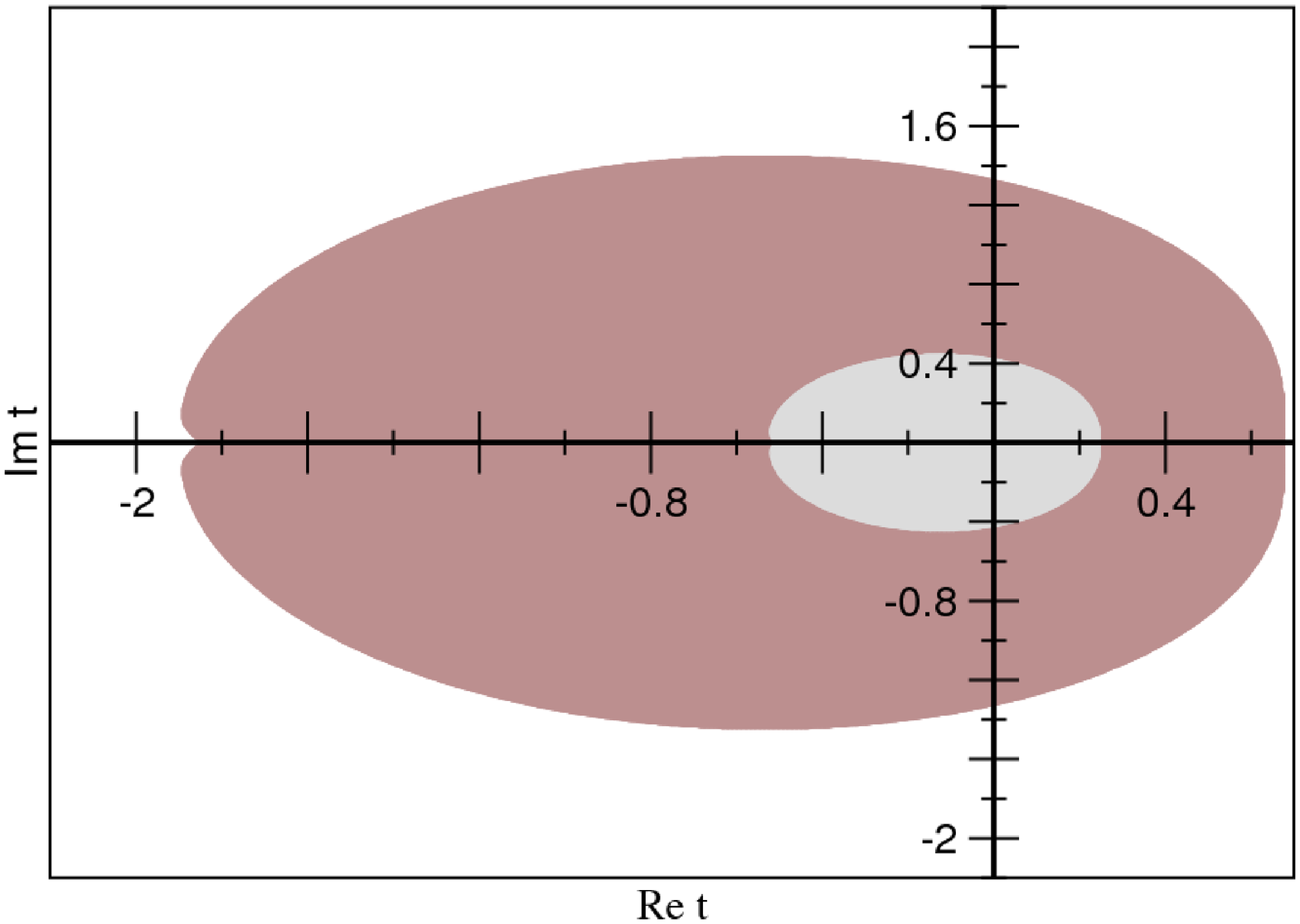}
\caption{Domain without zeros for the vector form factor: the small domain is obtained 
without including phase and modulus in the elastic region, bigger one using 
phase and modulus.}
\label{fig4}
\end{minipage}
\end{figure}
In Fig. \ref{fig1}, the allowed band for the slope $\lambda_0'$  is compared with the experimental determinations.  
The slope predicted 
by NA48 (2007) is not consistent with our predictions.  However their recent determination confirms our  predicted range, see 
Veltri et.al \cite{Veltri:2011zk}.  
We note that the theoretical prediction of ChPT to two loops 
$ \lambda_0'= (13.9_{+1.3}^{-0.4}\pm0.4)\times 10^{-3}$, $\lambda_0''= (8.0_{+0.3}^{-1.7})\times 10^{-4} $ 
is consistent within errors.  For the central value of the slope $ \lambda_0'$ given above, the range of  
$\lambda_0''$  is  $(8.24 \times
10^{-4}, 8.42\times 10^{-4})$. The same is true for the theoretical 
prediction $ \lambda_0'= (16.00 \pm 1.00 )\times 10^{-3}$, $\lambda_0''= (6.34\pm 0.38)\times 10^{-4} $ 
obtained from dispersion relations.

As shown in Fig. \ref{fig2} for vector form factor,  except the results from NA48 and KLOE, which have 
curvatures slightly larger than the allowed values, 
the experimental data satisfy the constraints.  
We note also that the theoretical predictions $ \lambda_+'= (24.9 \pm 1.3)\times 
10^{-3}$, $\lambda_+''= (1.6 \pm 0.5)\times 10^{-3} $  obtained from ChPT to two loops, and 
$ \lambda_+'= (26.05_{-0.51}^{+0.21})\times 10^{-3}$, $\lambda_+''= (1.29_{-0.04}^{+0.01})\times 10^{-3} $, 
and $ \lambda_+'= (25.49\pm0.31)\times 10^{-3}$, 
{\bf $\lambda_+''= (1.22\pm0.14)\times 10^{-3} $ } obtained from 
dispersion relations are consistent  with the constraint.
For precise results, see \cite{Abbas:2010ns}.  
As we mentioned, the same
formalism
can be used to derive regions in the complex plane where the form factors
can not vanish.
In Fig. \ref{fig3} we show the region where zeros of the scalar form factors are excluded.
If we  impose the Callan-Treiman constraint, the value of $f_0(\Delta_{K\pi})$,
the scalar form factor  cannot have simple zeros  in the range  
$-1.81  \, \mbox{GeV}^2 \leq t_0 \leq 0.93 \, \mbox{GeV}^2$.
The formalism  rules out zeros in the
physical region of the kaon semileptonic decay.
In the case of complex zeros, we have
obtained a rather large region where they cannot be present.

For the vector
form factors, simple zeros are excluded in the interval $-0.31\,
\mbox{GeV}^2 \leq  t_0 \leq 0.23\, \mbox{GeV}^2$ of the real axis, while
Fig. \ref{fig4} shows the region where complex zeros are excluded.
For more results, see \cite{Abbas:2010ns}. We mention that we do not use as input the soft-kaon theorem, 
but derive
bounds on the value at the relevant point. Thus we are able to predict a
narrow range $-0.046 \leq \bar{\Delta}_{CT} \leq 0.014$ for higher
order corrections.

\section{Conclusion}
We have studied  the shape of the scalar and vector
form factors in the $K_{\ell 3}$ domain,  crucial
for the determination of the modulus of the CKM matrix element $|V_{us}|$. 
The results are very stringent in the scalar form factor case. 
The most recent results from NA48 \cite{Veltri:2011zk} is consistent with our prediction for the slope of scalar form factor and
restricts the range of the slope to $\sim 0.01-0.02$.

Our results show that  the zeros are excluded in a rather large domain
at low energies, which provides confidence in semiphenomenological
analyses based on Omn\`es representations which assume  that the zeros
are absent. Unlike the standard dispersive treatments, our method does not
require  the knowledge of zeros and of the phase above the inelastic
threshold. Therefore, model dependent assumptions are not necessary.  The
price to pay is that we only are able to predict ranges for shape
parameters and zero positions.  However, due to the high precision of low
energy measurements and calculations, the predicted bounds are very
stringent.

\end{document}